\documentclass[a4paper,11pt]{article}
\pdfoutput=1
\usepackage{jcappub}
\usepackage[T1]{fontenc}
\allowdisplaybreaks[3]

\def\be{\begin{equation}}
\def\ee{\end{equation}}
\def\ba{\begin{eqnarray}}
\def\ea{\end{eqnarray}}
\def\ba{\begin{eqnarray}}
\def\ea{\end{eqnarray}}
\newcommand{\nn}{\nonumber\\}
\newcommand{\ud}{\mathrm{d}}
\newcommand{\uD}{\mathrm{D}}
\def \pd {\partial}
\def \bfx {{\bf x}}

\title{\boldmath Gradient expansion of superhorizon perturbations in G-inflation}

\author{Noemi Frusciante,}
\author{Shuang-Yong Zhou}
\author{and Thomas P. Sotiriou}

\affiliation{SISSA, Via Bonomea 265, 34136, Trieste, Italy and \\ 
INFN Sezione di Trieste, Italy}
\emailAdd{nfruscia@sissa.it}
\emailAdd{szhou@sissa.it}
\emailAdd{sotiriou@sissa.it} 

\abstract{We develop the gradient expansion formalism  for shift-symmetric Galileon-type actions. We focus on backgrounds that undergo inflation, work in the synchronous gauge, and obtain a general solution up to second order without imposing extra conditions at first order. The solution simplifies during the late stages of inflation. We also define a curvature perturbation conserved up to first order.}

\begin{document}
\keywords{Modified Gravity, Cosmological Perturbation Theory, Inflation, Non-Gaussianity.}


\maketitle
\flushbottom

\section{Introduction}

Inflation is a powerful paradigm that addresses various fine-tuning problems in the early Universe and accounts for the nearly scale invariant primordial perturbations that are needed for structure formation. These primordial perturbations leave an imprint on the cosmic microwave background (CMB). Single field slow-roll inflation models (with a canonical kinetic term) generically predict a Gaussian spectrum of primordial perturbations \cite{Maldacena:2002vr}. However, despite the success of the inflationary paradigm, its theoretical underpinning is still a matter of debate. Hence, it is not clear why one should remain within the framework of slow-roll inflation or even single-field inflation. 
Substantial non-Gaussianity can be generated in inflation models with multiple scalar fields or with non-canonical kinetic terms. Furthermore, if the slow-roll condition is temporarily violated, large non-Gaussianity can be generated even in a single field model  \cite{Chen:2006xjb}. On the observational side,  results from WMAP are consistent with a Gaussian spectrum of primordial perturbations \cite{Bennett:2012fp} and the recently released results from PLANCK are already leading to tighter constraints of non-Gaussianity \cite{Ade:2013tta}. With the prospect of probing the inflation scenario much deeper, non-Gaussianity in different inflation models has been extensively investigated and classified during the last decade (for a review, see \cite{Bartolo:2004if, Chen:2010xka}).

To tackle non-Gaussianity from inflation models, traditional linear perturbation theory is inadequate. A natural approach is to go beyond the linear order and work with second order cosmological perturbation theory \cite{Maldacena:2002vr, Acquaviva:2002ud, Malik:2003mv, Nakamura:2004rm, Malik:2005cy, Langlois:2006vv, Christopherson:2011hn}. While this approach usually applies to primordial perturbations up to the horizon exit, an alternative approach naturally tackles the superhorizon perturbations --- gradient expansion \cite{Lifshitz:1963ps,Tomita:1975kj,Salopek:1990jq,Comer:1994np, Deruelle:1994iz, Lyth:2004gb, Iguchi:1996jy, Tanaka:2006zp, Takamizu:2008ra, Izumi:2011eh, Gumrukcuoglu:2011ef,Takamizu:2013gy, Naruko:2012fe}.  In gradient expansion, physical quantities are expanded in terms of their inverse wavelengths, as compared to a pivotal length scale ($\epsilon \sim L_p/L_{phys}$), so every spatial derivative adds one perturbative order, $\pd_i\sim \epsilon$, hence th
 e name. This is different from usual cosmological perturbation theory where the expansion is in terms of perturbative field amplitudes. In the context of cosmology, particularly in the inflationary epoch when physical modes are stretched well outside the quasi-constant Hubble horizon,  the Hubble length can be naturally chosen as the pivotal length scale. Therefore, this approach can be used to evaluate and evolve non-Gaussiantites at superhorizon scales, complementary to usual non-linear perturbation theory.  The leading order gradient expansion is often called the separate universe approach \cite{Wands:2000dp} or $\delta N$ formalism \cite{Starobinsky:1986fxa, Sasaki:1995aw}, which is sufficient for many purposes. However,  the next-to-leading order gradient corrections can be as important, for example, in some multi-field models or when the slow-roll condition is violated \cite{Leach:2001zf, Naruko:2012fe}. A  beyond-$\delta N$ formalism scheme has recently been proposed~
 \cite{Naruko:2012fe}.

The Galileon is a scalar field with a galilean(-like) symmetry $\phi \to \phi + b_\mu x^\mu + c$ ($b_\mu , c$ being constant) around flat space, and was originally introduced as an effective, infrared gravitational modification which can lead to self-accellerating solutions \cite{Nicolis:2008in}. The Galileon Lagrangian contains higher order derivatives but nevertheless leads to second order equations of motion, thus avoiding Ostrogranski ghosts. 

Coupling the Galileon covariantly to gravity and insisting on the requirement that the scalar and the metric satisfy second order equations forces one has to abandon galilean symmetry \cite{Deffayet:2009wt}. Ordinary shift symmetry $\phi \to \phi + c$ can be retained \cite{Kobayashi:2010cm} or abandoned as well. In the latter case, covariant actions for generalized Galileons coupled to gravity have been constructed in Refs.~\cite{Deffayet:2009mn, Deffayet:2011gz}. The 4-dimensional version of the action given in Ref.~\cite{Deffayet:2011gz} has been shown \cite{Kobayashi:2011nu} to be equivalent to the most general action for a scalar coupled to gravity that leads to second order equations of motion, given by Horndeski in the 1970s \cite{Horndeski:1974wa}. The Galileon model has also been generalized in various other directions (see e.g.~\cite{deRham:2010eu, Padilla:2010tj, Deffayet:2010zh, Padilla:2010ir, Zhou:2011ix, Goon:2012dy, Cai:2012va, Hinterbichler:2011tt} and referen
 ces therein). 

The self-accelerating solutions of Galileon models have been the basis of inflationary scenarios \cite{Creminelli:2010ba, Kobayashi:2010cm, Burrage:2010cu, Creminelli:2010qf,  Kobayashi:2011nu, RenauxPetel:2011uk, Fasiello:2013dla, Kamada:2010qe, Kobayashi:2011pc, DeFelice:2013ar, Gao:2011qe}. There are some known novel features in such scenarios: to name a few, the null energy condition can be drastically violated without developing instabilities \cite{Creminelli:2010ba}; a large tensor-to-scalar power spectrum ratio is allowed \cite{Kobayashi:2010cm}; there are new shapes of the three-point function and potentially large four-point function \cite{Creminelli:2010qf}.

 In this paper, we develop the superhorizon gradient expansion formalism for G-inflation \cite{Kobayashi:2010cm}, up to second order in gradient expansion. We focus the subclass of  actions for  which the scalar enjoys shift symmetry, as they are closer to the original idea of the Galileon and significantly simpler. Additionally, our goal is to explore the phenomenology associated with the non-linear derivative interactions of the scalar. Abandoning shift symmetry leads, amongst other terms, to allowing a potential for the scalar, which can lead to similar phenomenology and, therefore, obscures the role of the Galileon-type terms. We derive the general solution for an Friedman--Lema\^ itre--Robertson--Walker (FLRW) background, identify the degrees of freedom in the model and define a curvature perturbation conserved up to  $\mathcal{O}(\epsilon)$. We also consider how our results simplify in the limit where the background becomes de Sitter spacetime. Given that the latter is 
 an attractor of G-inflation, this approximation provides a good description at least for the later stages of inflationary expansion.

During the preparation of this manuscript ref.~\cite{Takamizu:2013gy} appeared, which also develops the superhorizon gradient expansion formalism for G-inflation without assuming shift symmetry. However, there are major differences: On the technical side, we work in the synchronous gauge, while ref.~\cite{Takamizu:2013gy} prefers the uniform expansion gauge; On the more substantial side in ref.~\cite{Takamizu:2013gy} it is assumed that $\pd_t h_{ij}(t,x)$ (see eq. \ref{metricorder}) is $\mathcal{O}(\epsilon^2)$, while we do not impose such condition. In this respect our results are more general.

The paper is organized as follows. In Section \ref{ginflation}, we specify the model to investigate and write down the equations of motion and in Section \ref{decomposition} we perform $3+1$ decomposition of the equations of motion. In Section \ref{gradientexpansion}, we establish the gradient expansion orders of relevant quantities. The equations of motion are solved up to $\mathcal{O}(\epsilon^2)$ in Section \ref{generalsolution}, and the general solution is summarized in Section \ref{sec:solutionsum}. In Section \ref{latetime}, we simplify the general solution in the de Sitter limit. Note that in Section \ref{generalsolution}, Section \ref{latetime} and  Appendix \ref{sec:gradexpeom}, we mostly suppress the background quantities' order indication ${}^{(0)}$ to simplify the equations. We conclude and discuss some future perspectives in Section \ref{conclusion}.

\section{Inflation with a shift-symmetric Galileon}\label{ginflation}

Following the arguments in the Introduction, we consider the action
\be
S = \int \ud^4 x \sqrt{-g} \left( \frac{M^2_{pl}}{2}R + K(X) -G(X)\Box\phi \right),
\ee
where $M_{pl}$ is the Planck mass, $R$ is the Ricci scalar and  $K(X)$ and $G(X)$ are general functions of $X=-\pd_\mu \phi \pd^\mu \phi / 2$. This action was considered in Ref.~\cite{Deffayet:2010qz}. Even though it is not the most general action for shift-symmetric generalized Galileon models (without assuming specific forms for the unspecified functions), it is the simplest one which includes the characteristic ``Galileon interactions'', {\em i.e.}~terms that contain second order derivatives of $\phi$. Note that  for $G=0$ we recover k-inflation \cite{ArmendarizPicon:1999rj}, and the model with $K=X,G=\alpha X$ ($\alpha=$const.) is the cubic covariant Galileon \cite{Deffayet:2009wt}, which in turn is linked to the DGP model \cite{Luty:2003vm}. The equations of motion for the metric, to which we will refer as Einstein equations, are given by
\be
M_{pl}^2 G_{\mu\nu} = T^{\phi}_{\mu\nu} ,
\ee
with
\be
T^{\phi}_{\mu\nu} = (K_X - G_X \Box \phi) \pd_\mu \phi \pd_\nu \phi  - 2 \pd_{(\mu} G \pd_{\nu)} \phi +  g_{\mu\nu} (K + \pd_\sigma G \pd^\sigma \phi) ,
\ee
where a subscript $X$ denotes partial differentiation  with respect to $X$.
Note that the energy momentum tensor takes the form of an imperfect fluid \cite{Pujolas:2011he}, thus this model does not fall under the existing formalism for a perfect fluid \cite{Tanaka:2006zp}. Thanks to the shift symmetry, the equation of motion for $\phi$ can be given in terms of the current 
\be
J^\mu = (K_X - G_X\Box\phi)\pd^\mu \phi -G_X \pd^\mu X ,
\ee
as
\be
\nabla_\mu J^\mu = 0.
\ee
It is worth mentioning that the scalar equation of motion is implied by the Einstein equations, i.e., once the Einstein equations are satisfied, the scalar equation of motion is automatically satisfied. In fact, this applies to any covariant scalar-tensor system, as we show in Appendix~\ref{stsystem}.

\section{$3+1$ decomposition}\label{decomposition}

Now, we perform $3 + 1$ decomposition of the equations of motion. First, we decompose the metric according to the Arnowitt-Deser-Misner (ADM) prescription
\be
\ud s^2 = - N^2 \ud t^2 + \gamma_{ij} (\ud x^i + N^i \ud t) (\ud x^j + N^j \ud t) .
\ee 
To reduce redundant gauge degrees of freedom and simplify equations, we make use of a gauge condition:
\be
N = 1 , \qquad N^i 
= 0,
\ee
which implies
\begin{eqnarray} \label{sgmetric}
&& g_{tt}=-1, \,\,\,\,\, g_{ti}=0, \,\,\,\,\, g_{ij}=\gamma_{ij},\\ 
&& g^{tt}=-1, \,\,\,\,\, g^{ti}=0, \,\,\,\,\, g^{ij}=\gamma^{ij} .
\end{eqnarray}
Here latin indices (except for $t$) run from 1 to 3 and they are raised and lowered with $\gamma^{ij}$ and $\gamma_{ij}$ respectively. This is called synchronous gauge,  where the proper time distance between two neighboring hypersurfaces along the normal vector coincides with the coordinate time distance ($N=1$, proper time slicing) and the spatial coordinates are such that clocks are synchronized between different hypersurfaces ($N^i=0$). In synchronous gauge, equations can be very much simplified. Note, however, that there is residual gauge freedom, which will be discussed in section \ref{sec:solutionsum}.

Now, the spatial $\gamma_{ij}$ can be considered as a fundamental dynamical variable. 
 Another fundamental variable after the $3 + 1$ decomposition is the extrinsic curvature, which in synchronous gauge is simply
\begin{equation}
\mathcal{K}_{ij}=-\Gamma^t_{ij}=-\frac{1}{2}\dot{\gamma}_{ij},
\end{equation}
and its trace is defined as $\mathcal{K}=\gamma^{ij} \mathcal{K}_{ij}$. It is useful to decompose the spatial metric and the extrinsic curvature as
\begin{align}
\label{gammaijdef}
\gamma_{ij} &= a^2(t)e^{2\zeta(t,{\bfx})}h_{ij}(t,{\bfx}),
\\
\label{kijdef}
\mathcal{K}^i_j &= \frac{1}{3}\mathcal{K}(t,{\bfx})\delta^i_j+A^i_j(t,{\bfx}),
\end{align}
where $a(t)$ is the scale factor of the fiducial FLRW background, $\zeta(t,{\bfx})$ is related to the curvature perturbation, $h_{ij}(t,{\bfx})$ is defined to have a unit determinant $\det[h_{ij}]=1$,  and $A^i_j$ is the traceless part of $\mathcal{K}^i_j$.  These definitions lead to the following relations
\begin{eqnarray}
&&\mathcal{K}=-3\left[\frac{\dot{a}}{a}+\dot{\zeta}\right] ,  \label{Ktrace}
\\
&&\dot{h}_{ij}=-2h_{ik}A^k_j.\label{tderivmetric}
\end{eqnarray}

To decompose the equations of motion, we first make use of some well-known results which do not make reference to any specific gauge. Using the ADM variables, the unit normal 1-form and vector can be written respectively as $n_\mu =  (-N,0,0,0)$ and $n^\mu = (1/N, -N^1/N,-N^2/N,-N^3/N)$.  Making use of the Gauss-Codazzi relations (see e.g. \cite{Gourgoulhon:2007ue}), we can write the Ricci tensor and Ricci scalar respectively as
\begin{align}
R_{\mu\nu} &=   n_\mu n_\nu \left( \frac1N \mathcal{L}_\mathbf{m} \mathcal{K} + \frac1N \uD^\lambda \uD_{\lambda} N - \mathcal{K}^\rho_\sigma \mathcal{K}^\sigma_\rho \right) -  2n_{(\mu} \uD_{\nu)} \mathcal{K} + 2n_{(\mu|} \uD_{\sigma} \mathcal{K}^{\sigma}_{|\nu)} 
\nn
&~~~ - \frac1N \mathcal{L}_{\mathbf{m}} \mathcal{K}_{\mu\nu} - \frac1N \uD_\mu \uD_\nu N + {}^{[3]}R_{\mu\nu} + \mathcal{K} \mathcal{K}_{\mu\nu} - 2 \mathcal{K}_{\mu}^{\sigma} \mathcal{K}_{\nu{\sigma}},
\\
R &= {}^{[3]}R + \mathcal{K}^2 + \mathcal{K}^\rho_\sigma \mathcal{K}^\sigma_\rho - \frac2N \mathcal{L}_{\mathbf{m}}\mathcal{K} - \frac2N D^\sigma D_\sigma N,
\end{align}
where $m^\mu = N n^\mu$, $\mathcal{L}_{\mathbf{m}}$ is the Lie derivative along $m^\mu$, and $\uD_\mu$ is the covariant derivative, ${}^{[3]}R$ the Ricci scalar, and $ {}^{[3]}R_{\mu\nu}$ the Ricci tensor of the spacelike hypersurfaces. The  Laplacian is decomposed as
\be
\Box \phi = - n^\rho \pd_\rho (n^\sigma \pd_\sigma \phi) + \mathcal{K} n^\sigma \pd_\sigma \phi 
+ \uD^\sigma \ln N \pd_\sigma \phi + \uD^\sigma \uD_\sigma \phi .
\ee

In the synchronous gauge ($N=1, N^i=0$), the Einstein equations are greatly simplified:
\begin{align}
M_{pl}^2 G_{tt} &= T^\phi_{tt},
\\
M_{pl}^2 G_{ti} &= T^\phi_{ti} ,\label{tiequation}
\\
M_{pl}^2 G^i_j &= T^\phi|^i_j,
\end{align}
with
\begin{align}
G_{tt} &= \frac{1}{2}\left( {}^{[3]}R+ \mathcal{K}^2-\mathcal{K}^{i}_j \mathcal{K}^{j}_i  \right),
\\
G_{ti} &=  -\uD_k\mathcal{K}^k_i + \uD_i \mathcal{K},
\\
G^i_j &= {}^{[3]}G^i_{j}  -\dot{\mathcal{K}}^i_j  +\mathcal{K}\mathcal{K}^i_{j}  -\frac{1}{2}\delta^{i}_j \left( -2\dot{\mathcal{K}}+\mathcal{K}^2+\mathcal{K}^k_{l}\mathcal{K}^l_k \right),
\\
T^\phi_{tt} & = {K_X}\dot{\phi}^2 - {K} - {G_{X }}\Box\phi\dot{\phi}^2 - {G_{X }}\dot{\phi}\dot{X} - {G_{X }} \pd_k X \pd^k \phi,
\\
T^\phi_{ti} & = {K_X}\dot{\phi}\pd_i\phi - {G_{X }}\Box\phi\dot{\phi}\pd_i\phi - {G_{X }}\dot{X}\pd_i\phi - {G_{X }}\pd_i X\dot{\phi},
\\
T^\phi|^i_j & = K_X\pd^i\phi\pd_j\phi  - G_{X }\Box\phi\pd^i\phi\pd_j\phi - G_{X }\pd^iX\pd_j\phi  - G_{X }\pd^i\phi \pd_j X
\nn
& ~~~  + \left( G_{X }\pd^k\phi\pd_k X + K  - G_{X }\dot{X}\dot{\phi} \right)\delta^i_j,
\end{align}
where $\Box \phi = - \ddot{\phi} + \mathcal{K} \dot{\phi}  + \uD^\sigma \uD_\sigma \phi$. The scalar equation of motion is given by
\be  \label{scalarequation}
\pd_\mu  J^\mu + \frac12 \pd_\mu \ln{\gamma}  J^\mu =  \dot{J}^t  + (3H +3\dot{\zeta}) J^t  +  3\pd_i \zeta J^i + \pd_i  J^i  = 0,
\ee
where $\gamma = \det[\gamma_{ij}]$ and $H=\dot{a}/a$ is the usual Hubble parameter.

\section{Gradient expansion: order analysis}\label{gradientexpansion}

In standard cosmological perturbation theory one expands perturbatively in the field amplitudes.  To tackle non-Gaussianities in inflation models, second order perturbation theory is often used within the Hubble horizon. However, for physics processes at superhorizon scales one usually resorts to the gradient expansion technique. Note that the separate universe approach or the $\delta N$ formalism is simply the leading order gradient expansion \cite{Rigopoulos:2003ak}. Assuming  the characteristic spatial length is   $L_{phys}$, the dimensionless perturbative expansion parameter is 
\be
\epsilon\sim \frac{H^{-1}}{L_{phys}} \ll 1.
\ee
This means in particular that every spatial partial derivative carries an order of $\epsilon$
\be
\pd_i \sim \mathcal{O}(\epsilon),
\ee
while the time derivative is considered $\mathcal{O}(\epsilon^0)$.  The superhorizon gradient expansion is complimentary to the usual non-linear cosmological perturbation analysis and may capture fully nonlinear (in terms of the field amplitudes) physics at superhorizon scales, while the equations are still tractable due to the perturbative approach.

To perform the superhorizon perturbation analysis, we first need to deduce the starting orders for various quantities of interest. First, note that the equations of motion at $\mathcal{O}(\epsilon^0)$  should  simply determine the evolution of the scale factor $a(t)$ and the scalar, as the spacetime is supposed to be described by an FLRW line element.  Given the definition (\ref{gammaijdef}), one can infer that $h_{ij}$ should start with  $h^{(0)}_{ij}({\bfx})$; otherwise the $\mathcal{O}(\epsilon^0)$ equation would pick up terms involving $\pd_t h^{(0)}_{ij}(t,{\bfx})$, which is non-FLRW. From the scalar's equation of motion eq. (\ref{scalarequation}), we can infer that at $\mathcal{O}(\epsilon^0)$ the scalar field should be spatially homogeneous, meaning that $\phi$ starts with $\phi^{(0)}(t)$. Unlike previous work on this subject (see e.g.~\cite{Takamizu:2013gy}), we do not impose any conditions on the higher orders of these quantities. Therefore, we have 
\begin{align} 
{\rm starting~order~of~}\dot{h}_{ij} &= \mathcal{O}(\epsilon),\label{metricorder}
\\
{\rm starting~order~of~}\pd_i\phi &= \mathcal{O}(\epsilon^2). \label{phiassume}
\end{align}
Expanding eq.~(\ref{tderivmetric}) perturbatively (for $n\geq 1$)
\begin{equation}
\dot{h}_{ij}^{(n)}=-2\sum^{n-1}_{p=0}h^{(p)}_{ik}\left(A^{(n-p)}\right)^k_j 
\end{equation}
and making use of eq.~(\ref{metricorder}),  we can infer that 
\be
{\rm starting~order~of~}A^k_j=\mathcal{O}(\epsilon) .
\ee
Expanding eq.~(\ref{tiequation}), we infer that
\be
\partial_i\mathcal{K}^{(0)} =  0 .
\ee
Therefore, $\mathcal{K}^{(0)}$ is a function of $t$. From the definition (\ref{Ktrace}), and given that one can always redefine the scalar factor $a(t)$ to absorb $\zeta^{(0)}(t)$, it follows that 
\begin{align}
{\rm starting~order~of~} \dot{\zeta}&=\mathcal{O}(\epsilon),
\\
\mathcal{K}^{(0)}&=-3\frac{\dot{a}}{a}=-3H(t),\\
\label{ntrace}
\mathcal{K}^{(n)}&=-3\dot{\zeta}^{(n)}, \qquad n \geq 1  ,
\end{align}
where $H(t)$ is the usual Hubble parameter.

Using eq.~(\ref{phiassume}), we may expand $X$ as
\begin{equation} \label{Xexpand}
X=X^{(0)}(t,{\bfx})+ X^{(1)}(t,{\bfx})\epsilon+ X^{(2)}(t,{\bfx})\epsilon^2+ \mathcal{O}(\epsilon^3) ,
\end{equation}
where
\begin{equation}
X^{(n)}=\frac{1}{2}\sum^{n}_{p=0}\dot{\phi}^{(p)}\dot{\phi}^{(n-p)}+\mathcal{O}(\epsilon^4).
\end{equation}
We also need to perturbatively expand functions of $X$, such as $K(X)$. To this end, we should consider $X=X(\epsilon)$ according to eq. (\ref{Xexpand}) and Taylor-expand, for example, $K(X(\epsilon))$ around $\epsilon=0$ as
\be
K(X)=K(X^{(0)}) +K_X (X^{(0)}) X^{(1)}\epsilon+\frac{1}{2}\left[K_{XX}(X^{(0)}) X^{(1)2}+2K_X(X^{(0)})  X^{(2)}\right]\epsilon^2+ \mathcal{O}(\epsilon^3).
\ee

In summary, the various quantities of interest, to be determined in the next section, are expanded as follows:
\begin{align}
\zeta &= \zeta^{(0)}({\bfx})+\zeta^{(1)}(t,{\bfx})\epsilon +\zeta^{(2)}(t,{\bfx})\epsilon^2 +\mathcal{O}(\epsilon^3),
\\
\phi &= \phi^{(0)}(t)+\phi^{(1)}(t,{\bfx})\epsilon + \phi^{(2)}(t,{\bfx})\epsilon^2+\mathcal{O}(\epsilon^3)\, ,  
\\
A^i_j &= A^{(1)}{}^{i}_j(t,{\bfx})\epsilon+ A^{(2)}{}^{i}_j(t,{\bfx})\epsilon^2+\mathcal{O}(\epsilon^3),
\\
h_{ij} & = h^{(0)}_{ij}({\bfx})+ h^{(1)}_{ij}(t,{\bfx})\epsilon+ h^{(2)}_{ij}(t,{\bfx})\epsilon^2+ \mathcal{O}(\epsilon^3)\, , 
\\
\mathcal{K}^i_j &= -H(t)\delta^i_j+\mathcal{K}^{(1)}{}^i_j(t,{\bfx})\epsilon + \mathcal{K}^{(2)}{}^i_j(t,{\bfx})\epsilon^2+\mathcal{O}(\epsilon^3),
\\
\mathcal{K} &= -3H(t)+\dot{\zeta}^{(1)}(t,{\bfx})\epsilon + \dot{\zeta}^{(2)}(t,{\bfx})\epsilon^2+\mathcal{O}(\epsilon^3).
\end{align}

\section{General solution}\label{generalsolution}

Now, we solve the equations of motion perturbatively, up to the $\mathcal{O}(\epsilon^2)$ order, to obtain the general solutions. These solutions will be parametrized by a few unspecified spatial functions, which describe the physical degrees of freedom (modulo residual gauge freedom) that may evolve as the Universe expands. The gradient expansions of Einstein's equations up to order $\mathcal{O}(\epsilon^2)$  are listed in Appendix~\ref{sec:gradexpeom}.

For the rest of this paper,  to simplify the equations, we will mostly suppress the background quantities' order indication ${}^{(0)}$. For example, $\dot{\phi}^{(0)}$ is written as $\dot{\phi}$ if there is no confusion.

\subsection{The $\mathcal{O}(\epsilon^0)$ order}

For the $\mathcal{O}(\epsilon^0)$ order, all spatial derivatives are absent. As desired, the equations of motion reduced to the conventional background FLRW case:
\begin{align}
3M^2_{pl}H^2 &= -K + 2K_X X +6H\dot{\phi}G_{X } X,\\
-M^2_{pl}\left(2\dot{H}+3H^2\right) &= K - 2G_{X } X \ddot{\phi},\\
\label{jteq}
\dot{J}^t{}^{(0)}+3HJ^t{}^{(0)} &=0, 
\end{align}
where  $J^t{}^{(0)}=K_X^{ }\dot{\phi} +6HG_{X } X$. Note that only two of the three equations are independent.

\subsection{The $\mathcal{O}(\epsilon)$ order}

The $tt$ component of Einstein's equation is given by
\be  \label{eett1order}
\left( \frac12 \dot{\phi} K_X + \dot{\phi} K_{XX} X + 9HG_X X + 6HG_{XX} X^2 \right) \dot{\phi}^{(1)}
=
\left(3M_{pl}^2 H - 3\dot{\phi} G_X X\right) \dot{\zeta}^{(1)} ,
\ee
which can be re-written as
\be \label{eett1}
\dot{\phi}^{(1)} = \mathcal{A}^0 \dot{\zeta}^{(1)} ,
\ee
where
\be \label{A0def}
\mathcal{A}^0(t) = \frac{6M_{pl}^2 H - 6\dot{\phi} G_X X}{ \dot{\phi} K_X + 2\dot{\phi} K_{XX} X + 18HG_X X + 12HG_{XX} X^2}.
\ee
The $ij$ component of Einstein's equation naturally splits into a trace part and a traceless part. The trace part gives rise to another relation between $\phi^{(1)}$ and $\zeta^{(1)}$:
\be \label{eeij1}
-2G_X X\ddot{\phi}^{(1)} + \left( \dot{\phi}K_X -  2\dot{X} (G_X + G_{XX} X) \right)\dot{\phi}^{(1)}    + 2M_{pl}^2\left(\ddot{\zeta}^{(1)}+3H\dot{\zeta}^{(1)}\right) = 0 .
\ee
Combining eq. (\ref{eett1}) and eq. (\ref{eeij1}), we get, after integration, 
\be
 {\zeta}^{(1)}(t,\bfx)  = {C}^{(1)}_\zeta(\bfx) \int^t \frac{\ud t'}{\bar{a}(t')^3} ,
 \ee
where 
\begin{align} \label{abardef}
\bar{a}(t) &= \exp(\int^t \ud t' \mathcal{H}^0(t') ) ,
\\
\mathcal{H}^0(t) &= \frac{\left( \dot{\phi}K_X -  2\dot{X} (G_X + G_{XX} X) \right) \mathcal{A}^0  -2G_X X \dot{\mathcal{A}}^0  + 6M_{pl}^2H}{6M_{pl}^2-6G_X X {\mathcal{A}}^0} ,
\end{align}
and
$ \mathcal{C}^{(1)}_\zeta(\bfx)$ is an unspecified spatial function from the first integration. There would be another unspecified spatial function from the second integration (${C}'^{(1)}_\zeta(\bfx)$), which has been absorbed into $\zeta^{(0)}(\bfx)$. We will see in the next section that $\mathcal{H}^0$ approaches the Hubble constant $H$ for near de Sitter expansion, in which case $\dot{\zeta}^{(1)}$ scales as $1/a^3(t)$.  Now,   $\phi^{(1)}$ is given by
\be
\phi^{(1)}(t,\bfx) =  {C}^{(1)}_\zeta(\bfx)\int^t  \frac{\ud t' \mathcal{A}^0(t')}{\bar{a}(t')^3}  + C^{(1)}_\phi(\bfx)  ,
\ee
where $C^{(1)}_\phi(\bfx)$ is an integration spatial function. The traceless part of Einstein's equation's $ij$ component is simply
\be
\dot{A}^{(1)}{}^{i}_{j}+3HA^{(1)}{}^{i}_{j} = 0 ,
\ee
whose solution is
\begin{equation}
A^{(1)}{}^i_{j}(t,{\bfx}) = \frac{{C}^{(1)}_{A}{}^i_j({\bfx})}{a^3} , 
\end{equation}
where the unspecified spatial function ${C}^{(1)}_{A}{}^i_j({\bfx})$ is symmetric and traceless. From eq. (\ref{tderivmetric}), we have
\begin{equation}
h^{(1)}_{ij}(t,{\bfx})=-2h^{(0)}_{ik}({\bfx}) C^{(1)}_{A}{}^{k}_{j}({\bfx}) \int^t  \frac{\ud t'}{a(t')^3} ,
\end{equation}
where the would-be integration spatial function $C^{(1)}_{h}{}_{ij}({\bfx})$ has been absorbed into $h^{(0)}_{ik}({\bfx})$. As expected, the scalar equation of motion is solved by the solution obtained above.

Before moving on to solve higher order equations, we note that defining a curvature perturbation that is conserved in time is trivial in our formalism. By  virtue of the $tt$ component of Einstein's equation (\ref{eett1}), one can define a conserved curvature perturbation at $\mathcal{O}(\epsilon)$
\be
\mathcal{R}^{(1)} =  {\zeta}^{(1)} - \int^t  \frac{\ud t'}{\mathcal{A}^0(t')} \dot{\phi}^{(1)}(t') .
\ee
As we will see in Section~\ref{latetime}, because of the shift symmetry, de Sitter expansion is an attractor of the system.  For quasi-de Sitter expansion, i.e., for the late time of inflation,  $\mathcal{A}^0\simeq {\rm constant}$ and we can write $\mathcal{R}^{(1)} $ as
\be
\mathcal{R}^{(1)}  \simeq {\zeta}^{(1)} -\frac{1}{\mathcal{A}^0} {\phi}^{(1)} .
\ee

\subsection{The $\mathcal{O}(\epsilon^2)$ order}

The $tt$ component of the Einstein equation gives 
\begin{eqnarray} \label{phizeta2eq}
\mathcal{A}^0{}\dot{\zeta}^{(2)}-\dot{\phi}^{(2)}=\mathcal{C}^0 \left(\dot{\zeta}^{(1)}\right)^2-\frac{\mathcal{C}^0_3}{2}\,^{[3]}R^{(2)}+\frac{\mathcal{C}^0_3}{2}A^{(1)}{}^k_{j}A^{(1)}{}^j_{k},
\end{eqnarray}
where $\,^{[3]}R^{(2)}$ is the 3D Ricci scalar for the $\mathcal{O}(\epsilon^0)$ order metric $\gamma_{ij} = a(t)^2 e^{2\zeta^{(0)}(\bfx)} h^{(0)}_{ij}(\bfx)$ and $\mathcal{C}^0$, $\mathcal{C}^0_1$, $\mathcal{C}^0_2$ and $\mathcal{C}^0_3$ are again background quantities, defined respectively as
\begin{align}
\label{Cdef}
\mathcal{C}^0(t)&={\mathcal{C}^0_1(\mathcal{A}^0)^2}+{\mathcal{C}^0_2\mathcal{A}^0{}}-{3\mathcal{C}^0_3},
\\
\label{Cdef1}
\mathcal{C}^0_1(t)&=\frac{\frac12 K_X + 4K_{XX}X + 2 K_{XXX}X^2 + 9 H\dot{\phi} G_{X} + 21 H\dot{\phi} G_{XX} X + 6 H\dot{\phi}G_{XXX}X^2}{\dot{\phi}K_{X}+2\dot{\phi}K_{XX}X +18 HG_X X + 12 H G_{XX} X^2},
\\
\label{Cdef2}
\mathcal{C}^0_2(t)&=\frac{18G_X X + 12 G_{XX} X^2}{\dot{\phi}K_{X}+2\dot{\phi}K_{XX}X +18 HG_X X + 12 H G_{XX} X^2} , 
\\
\label{Cdef3}
\mathcal{C}^0_3(t)&=\frac{M^2_{pl}}{\dot{\phi}K_{X}+2\dot{\phi}K_{XX}X +18 HG_X X + 12 H G_{XX} X^2}. 
\end{align}
Integrating this equation, we get the solution of $\phi^{(2)}$ in terms of $\zeta^{(2)}$:
\begin{align}
{\phi}^{(2)}(t,\bfx) &=\int^t \ud t'   \mathcal{A}^0(t')\dot{\zeta}^{(2)}(t',\bfx) - \left(\mathcal{C}^{(1)}_\zeta(\bfx)\right)^2 \int^t \frac{\ud t' \mathcal{C}^0(t')}{\bar{a}(t')^6} +\frac{\,{}^{[3]}R^{(2)}(\bfx)}{2} \int^t\frac{\ud t' \mathcal{C}^0_3(t')}{a(t')^2} 
\nn
&~~~ -\frac{C^{(1)}_A{}^k_j(\bfx)C^{(1)}_A{}^j_k(\bfx)}{2} \int^t \frac{\ud t'\mathcal{C}^0_3(t')}{a(t')^6},
\end{align}
where an integration spatial function has been absorbed into $C^{(1)}_\phi(\bfx)$, and $\,^{[3]}R^{(2)}(\bfx)$ (the Ricci scalar of the metric $e^{2\zeta^{(0)}(\bfx)} h^{(0)}_{ij}(\bfx)$)  is related to $\,^{[3]}R^{(2)}$ (the Ricci scalar of the metric $a(t)^2 e^{2\zeta^{(0)}(\bfx)} h^{(0)}_{ij}(\bfx)$) by 
\be
\,^{[3]}R^{(2)}(\bfx)= a(t)^2\,^{[3]}R^{(2)} .
\ee
The trace part of Einstein's equation's $ij$ component is given by
\begin{align}
&~~~ -M_{pl}^2\left( 2\ddot{\zeta}^{(2)}+6H\dot{\zeta}^{(2)}+3\left(\dot{\zeta}^{(1)}\right)^{2}+\frac{1}{2}A^{(1)}{}^{k}_{l}A^{(1)}{}^{l}_{k} +  \frac16 \,^{[3]}R^{(2)}\right) 
\nn
&=  -2 G_X X \ddot{\phi}^{(2)} + \left(\dot{\phi} K_X - 2 \dot{X}(G_X + G_{XX}X) \right) \dot{\phi}^{(2)}  +\mathcal{D}^0 \left(\dot{\phi}^{(1)}\right)^2 ,
\end{align}
where
\begin{align}
\mathcal{D}^0 (t) &= \Big( \frac12 K_X +K_{XX} X - \ddot{\phi} G_X  - 5\ddot{\phi} G_{XX}X - 2\ddot{\phi} G_{XXX}X^2  
\nn
&\qquad~~ + 2 \left(3\mathcal{H}^0- \pd_t\ln \mathcal{A}^0 \right)\dot{\phi} (G_X + G_{XX} X) \Big) .
\end{align}
Combining with eq. (\ref{phizeta2eq}), we get
\begin{align}
\zeta^{(2)}(t,{\bfx}) &= \left(C^{(1)}_{\zeta}(\bfx)\right)^2   \int^t \frac{ \ud t''}{\bar{a}(t'')^3} \int^{t''} \frac{ \ud t' \mathcal{E}_1^0 (t')}{\bar{a}(t')^3} 
+  {}^{[3]}R^{(2)}(\bfx) \int^t \frac{ \ud t''}{\bar{a}(t'')^3} \int^{t''} \frac{ \ud t' \mathcal{E}_3^0 (t') \bar{a}(t')^3}{a(t')^2}
\nn
&~~~   + C_A^{(1)}{}^k_l({\bfx})C^{(1)}_A{}^l_k({\bfx})  \int^t \frac{ \ud t''}{\bar{a}(t'')^3} \int^{t''} \frac{ \ud t' \mathcal{E}_2^0 (t') \bar{a}(t')^3}{a(t')^6}    ,   
\end{align}
where two integration spatial functions have been absorbed into $C^{(1)}_{\zeta}({\bfx})$ and $\zeta^{(0)}(\bfx)$ respectively, and $\mathcal{E}^0_1$, $\mathcal{E}^0_2$ and $\mathcal{E}^0_3$ are background quantities, defined respectively as
\begin{align}
\label{Edef1}
\mathcal{E}_1^0 (t)&= \frac{ \left(\dot{\phi} K_X - 2 \dot{X}(G_X + G_{XX}X) \right) \mathcal{C}^0 + 2G_XX(6\mathcal{H}^0 \mathcal{C}^0 - \dot{\mathcal{C}}^0) - \mathcal{D}^0\left(\mathcal{A}^0\right)^2 - 3M^2_{pl}}{2M^2_{pl}-2G_X X \mathcal{A}^0},
\\
\label{Edef2}
\mathcal{E}_2^0 (t)&= \frac{\left(\dot{\phi} K_X - 2 \dot{X}(G_X + G_{XX}X) \right)\mathcal{C}^0_3 + 2G_XX(6H \mathcal{C}^0_3 - \dot{\mathcal{C}}^0_3 ) - M_{pl}^2}{4M^2_{pl}-4G_X X \mathcal{A}^0},
\\
\label{Edef3}
\mathcal{E}_3^0 (t)&= - \frac{\left(\dot{\phi} K_X - 2 \dot{X}(G_X + G_{XX}X) \right)\mathcal{C}^0_3 + 2G_XX(2H \mathcal{C}^0_3 - \dot{\mathcal{C}}^0_3 ) + \frac13 M_{pl}^2}{4M^2_{pl}-4G_X X \mathcal{A}^0}.
\end{align}
The traceless part of Einstein's equation's $ij$ component is given by
\be
 \dot{A}^{(2)}{}^{i}_{j}+3HA^{(2)}{}^{i}_{j}+3\dot{\zeta}^{(1)}A^{(1)}{}^{i}_{j}- \left(\,^{[3]}R^{(2)}{}^{i}_{j} -\frac13 \delta^i_j  \,^{[3]}R^{(2)} \right)  =0,
\ee
which gives rise to the solution
\be
A^{(2)}{}^{i}_{j}(t,{\bfx})= -\frac{3C^{(1)}_{\zeta}(\bfx){C}^{(1)}_{A}{}^i_j({\bfx})}{a^3}\int^t \frac{ \ud t'}{\bar{a}(t')^3}
+\frac{^{[3]}R^{(2)}{}^i_j(\bfx) -\frac13 \delta^i_j {}^{[3]}R^{(2)}(\bfx)}{a^3}\int^t \ud t' a(t') , 
\ee
where again an integration spatial function has been absorbed into $C^{(1)}_{A}{}^k_j({\bfx})$ and $\,^{[3]}R^{(2)}{}^i_j(\bfx)$ (the Ricci tensor of the metric $e^{2\zeta^{(0)}(\bfx)} h^{(0)}_{ij}(\bfx)$)  is related to $\,^{[3]}R^{(2)}{}^i_j$ (the Ricci tensor of the metric $a(t)^2 e^{2\zeta^{(0)}(\bfx)} h^{(0)}_{ij}(\bfx)$) by 
\be
\,^{[3]}R^{(2)}{}^i_j(\bfx)= a(t)^2\,^{[3]}R^{(2)}{}^i_j .
\ee
From eq. (\ref{tderivmetric}), we can derive 
\begin{align}
h^{(2)}_{ij}(t,\bfx)&=6 h^{(0)}_{ik}(\bfx) C^{(1)}_{\zeta}(\bfx){C}^{(1)}_{A}{}^k_j({\bfx}) \int^t \!\!\frac{\ud t''}{a(t'')^3}\int^{t''}\!\! \frac{ \ud t'}{\bar{a}(t')^3} 
\nn
&~~~ + 4 h^{(0)}_{il}(\bfx)C_A^{(1)}{}^l_k(\bfx)C_A^{(1)}{}^k_j(\bfx) \int^t \!\!\frac{\ud t''}{a(t'')^3}\int^{t''}\!\! \frac{ \ud t'}{{a}(t')^3}
\nn
&~~~  -2 h^{(0)}_{ik}(\bfx)  \left({}^{[3]}R^{(2)}{}^k_j(\bfx) -\frac13 \delta^k_j {}^{[3]}R^{(2)}(\bfx)\right)\int^t \!\!\frac{\ud t''}{a(t'')^3}\int^{t''}\!\! { \ud t' {a}(t')} ,
\end{align}
where  an integration spatial function has been absorbed into $h^{(0)}{}_{ij}(\bfx)$. The $ti$ component of Einstein's equation at the $\mathcal{O}(\epsilon^2)$ order become constraints for the $\mathcal{O}(\epsilon)$ order quantities
\begin{equation} \label{ti2order}
 -2M_{pl}^2\pd_i \dot{\zeta}^{(1)} - M_{pl}^2\uD^{(1)}_k A^{(1)}{}^k_i =(K_X^{ }\dot{\phi} + 6HG_{X } X) \pd_i \phi^{(1)} - 2G_X X \pd_i \dot{\phi}^{(1)},
\end{equation}
where  $\uD^{(1)}_k$, of order $\mathcal{O}(\epsilon)$ itself,  is the covariant derivative associated with the $\mathcal{O}(\epsilon^0)$ order metric $e^{2\zeta^{(0)}(\bfx)} h^{(0)}_{ij}(\bfx)$. This gives rise to 3 constraints on the unspecified integration functions ${C}^{(1)}_{A}{}^i_j({\bfx})$.

\subsection{Summary} \label{sec:solutionsum}

Here we summarize the solution obtained up to the $\mathcal{O}(\epsilon^2)$ order:
\begin{align}
{\zeta}(t,\bfx)  &= \zeta^{(0)}(\bfx) + {C}^{(1)}_\zeta(\bfx) \! \int^t \!\! \frac{\ud t'}{\bar{a}(t')^3}  + \left(C^{(1)}_{\zeta}(\bfx)\right)^2 \!\!  \int^t\!\! \frac{ \ud t''}{\bar{a}(t'')^3} \int^{t''}\!\! \frac{ \ud t' \mathcal{E}_1^0 (t')}{\bar{a}(t')^3} 
\nn
&~~~ + C_A^{(1)}{}^k_l({\bfx})C^{(1)}_A{}^l_k({\bfx})  \int^t \frac{ \ud t''}{\bar{a}(t'')^3} \int^{t''} \frac{ \ud t' \mathcal{E}_2^0 (t') \bar{a}(t')^3}{a(t')^6}  
\nn
&~~~ +  {}^{[3]}R^{(2)}(\bfx)\! \int^t\!\! \frac{ \ud t''}{\bar{a}(t'')^3} \!\int^{t''}\!\! \frac{ \ud t' \mathcal{E}_3^0 (t') \bar{a}(t')^3}{a(t')^2}    
+\mathcal{O}(\epsilon^3),
\\
\phi(t,\bfx) &= \phi^{(0)}(t) + C^{(1)}_\phi(\bfx)+ {C}^{(1)}_\zeta(\bfx)\int^t  \frac{\ud t' \mathcal{A}^0(t')}{\bar{a}(t')^3}  +\int^t \ud t'   \mathcal{A}^0(t')\dot{\zeta}^{(2)}(t',\bfx) 
\nn
&~~~ - \left({C}^{(1)}_\zeta(\bfx)\right)^2 \int^t \frac{\ud t' \mathcal{C}^0(t')}{\bar{a}(t')^6} +\frac{\,{}^{[3]}R^{(2)}(\bfx)}{2} \int^t\frac{\ud t' \mathcal{C}^0_3(t')}{a(t')^2} 
\nn
&~~~ -\frac{C^{(1)}_A{}^k_j(\bfx)C^{(1)}_A{}^j_k(\bfx)}{2} \int^t \frac{\ud t'\mathcal{C}^0_3(t')}{a(t')^6}
+\mathcal{O}(\epsilon^3),
\\
A{}^i_{j}(t,{\bfx}) &= \frac{{C}^{(1)}_{A}{}^i_j({\bfx})}{a^3}   -\frac{3C^{(1)}_{\zeta}(\bfx){C}^{(1)}_{A}{}^i_j({\bfx})}{a^3}\int^t \frac{ \ud t'}{\bar{a}(t')^3}
\nn
&~~~ +\frac{^{[3]}R^{(2)}{}^i_j(\bfx) -\frac13 \delta^i_j {}^{[3]}R^{(2)}(\bfx)}{a^3}\int^t \ud t' a(t')
+\mathcal{O}(\epsilon^3),
\\
h_{ij}(t,{\bfx}) &= h^{(0)}_{ij}(\bfx) -2h^{(0)}_{ik}({\bfx}) C^{(1)}_{A}{}^{k}_{j}({\bfx}) \int^t\!\!  \frac{\ud t'}{a(t')^3}
\nn
&~~~  +6 h^{(0)}_{ik}(\bfx) C^{(1)}_{\zeta}(\bfx){C}^{(1)}_{A}{}^k_j({\bfx}) \int^t \!\!\frac{\ud t''}{a(t'')^3}\int^{t''}\!\! \frac{ \ud t'}{\bar{a}(t')^3} 
\nn
&~~~ + 4 h^{(0)}_{il}(\bfx)C_A^{(1)}{}^l_k(\bfx)C_A^{(1)}{}^k_j(\bfx) \int^t \!\!\frac{\ud t''}{a(t'')^3}\int^{t''}\!\! \frac{ \ud t'}{{a}(t')^3}
\nn
&~~~  -2 h^{(0)}_{ik}(\bfx)  \left({}^{[3]}R^{(2)}{}^k_j(\bfx) -\frac13 \delta^k_j {}^{[3]}R^{(2)}(\bfx)\right)\int^t \!\!\frac{\ud t''}{a(t'')^3}\int^{t''}\!\! { \ud t' {a}(t')} +\mathcal{O}(\epsilon^3),
\end{align} 
where $\mathcal{A}^0$ is defined by eq. (\ref{A0def}), $\bar{a}(t)$ is defined by eq. (\ref{abardef}), $\mathcal{C}^0$, $\mathcal{C}^0_1$, $\mathcal{C}^0_2$ and $\mathcal{C}^0_3$ are defined by eqs. (\ref{Cdef}-\ref{Cdef3}) respectively,  $\mathcal{E}^0_1$, $\mathcal{E}^0_2$ and $\mathcal{E}^0_3$ are defined by eqs. (\ref{Edef1}-\ref{Edef3}) respectively, ${}^{[3]}R^{(2)}{}^i_j(\bfx)$ and ${}^{[3]}R^{(2)}(\bfx)$ are 3D curvature tensors of the metric $e^{2\zeta^{(0)}(\bfx)} h^{(0)}_{ij}(\bfx)$.

There are several unspecified spatial functions in the general solution: $\zeta^{(0)}(\bfx)$, $h^{(0)}_{ij}(\bfx)$, $C^{(1)}_{\zeta}(\bfx)$, $C^{(1)}_{\phi}(\bfx)$ and $C^{(1)}_A{}^{i}_j(\bfx)$ (all other unspecified spatial functions have been absorbed into this set of functions). These functions play the role of initial data for the dynamical degrees of freedom, so counting the pieces of  initial data and taking into account any constraints between them  can be used in order to determine the number of degrees of freedom. However, properly counting the physical degrees of freedom requires determining whether there are any degrees of freedom that can be removed using residual gauge freedom.

$h^{(0)}_{ij}(\bfx)$ is symmetric and has a unit determinant and $C^{(1)}_A{}^{i}_j(\bfx)$ is symmetric and traceless, so they each have 5 degrees of freedom. 3 components of $C^{(1)}_A{}^{i}_j(\bfx)$ are related to other unspecified spatial functions respectively by the constraint equations (\ref{ti2order}). In order to determine how many of these degrees of freedom are pure gauge we need to consider 
 the residual gauge freedom. Performing the  infinitesimal coordinate transformation
\be
x^\mu \to \bar{x}^\mu= x^\mu + \eta^\mu
\ee
and requiring that the synchronous gauge condition on the lapse $N$ and shift $N^i$ are respected, one straighforwadly obtains that $\eta^\mu$ should be of the form
\begin{align}
\label{lgeta0}
\eta^0&=\eta^0(\bfx),
\\
\label{lgetai}
\eta^i &= \int^t \ud t' \gamma^{ij}({t'},\bfx) \pd_j \eta^0(\bfx) + \tilde{\eta}^i(\bfx),
\end{align}
($\eta^\mu$ may be chosen as $\mathcal{O}(\epsilon)$). From this we infer that the residual gauge freedom amounts to 4 functions of space. 3 of those can be chosen so as to eliminate 3 spatial functions in $h^{(0)}_{ij}(\bfx)$ and 1 chosen so as to eliminate $C^{(1)}_\phi(\bfx)$.  Therefore, we may count the degrees of freedoms as follows:
\begin{align}
\zeta^{(0)}(\bfx) &\qquad{\rm 1~scalar~growing~mode~=~1~component},
\\
h^{(0)}_{ij}(\bfx)&\qquad{\rm 2~tensor~growing~modes~=~5~components~-~3~gauge~DoFs},
\\
C^{(1)}_{\zeta}(\bfx)&\qquad{\rm 1~scalar~decaying~mode~=~1~component},
\\
C^{(1)}_A{}^{i}_j(\bfx) &\qquad{\rm 2~tensor~decaying~modes~=~5~components~-~3~constraints}.
\end{align}
In this scalar-tensor system, one expects three physical degrees of freedom, one for the scalar mode and two for the tensor modes. As it is a second order system, each physical degree of freedom contains two phase-space degrees of freedom, so one should expect six free spatial functions. This is indeed what the counting reveals. 

\section{Late time of inflation} \label{latetime}

At the end of the last section, we claimed that $C^{(1)}_{\zeta}(\bfx)$ represents a decaying mode. However, this is actually not obvious from the general solution given above. After all, $\bar{a}(t)$ is not the scale factor $a(t)$ but is given by a rather complicated expression in terms of background quantities. Additionally, $\mathcal{C}^0_{n}$ and $\mathcal{E}^0_n$ also have time dependence. In this section, we would like to briefly re-derive the solution for an important special case, the late time of inflation. This will not only allows us to show explicitly that $C^{(1)}_{\zeta}(\bfx)$ represents a decaying mode, but it will demonstrate how one can eliminate the gauge mode $C^{(1)}_\phi(\bfx)$ on inflationary backgrounds. Moreover, the assumption of quasi-de Sitter expansion drastically simplifies the solution and allows an intuitive understanding of its behaviour. Physically, perturbations coming from the late time of inflation are observationally most important, as it
  is these perturbations that seed the large scale structure of the observable Universe.  Note that, similar to the previous section,  we mostly suppress the background quantities' order indication ${}^{(0)}$ to simplify the equations.

Eq. (\ref{jteq}) can be integrated to get
\begin{equation}
K_X^{ }\dot{\phi}^{ }+6HG_{X } X^{ }\propto a(t)^{-3},
\end{equation}
which is an attractor of the dynamical system. So for the later time of inflation $J^t$ essentially vanishes. In this limit, the background equations of motion can be simplified to
\begin{align}
K &=-3M^2_{pl}H^2,\\
K_X &=-3G_{X } H\dot{\phi}^{ },
\end{align}
and $\dot{\phi}$ and $H$ become constant and $a\propto e^{Ht}$ \cite{Kobayashi:2011nu}. Furthermore, we have 
\be
\mathcal{H}^0 \to H   , \qquad     \bar{a}(t) \to a(t)  .
\ee
The background quantities defined in the last section now all become constant and can also be simplified:
\begin{align}
\mathcal{A}^0{} &= \frac{3M^2_{pl}H-3\dot{\phi}G_{X } X}{\dot{\phi}^{ }K_{XX} X+6HG_{X } X+6HG_{X X} X^2},
\\
\mathcal{C}^0&={\mathcal{C}^0_1(\mathcal{A}^0)^2}+{\mathcal{C}^0_2\mathcal{A}^0{}}-{3\mathcal{C}^0_3},
\\
\mathcal{C}^0_1&=\frac{-\frac{5}{2}K_X+4K_{XX} X+2K_{XXX} X^2+21H\dot{\phi}G_{X X} X+6H\dot{\phi}G_{X XX} X^2}{2\dot{\phi}^{ }K_{XX} X+12HG_{X } X+12HG_{X X} X^2}, 
 \\
\mathcal{C}^0_2&=\frac{9G_{X } X+6G_{X X} X^2}{\dot{\phi}^{ }K_{XX} X+6HG_{X } X+6HG_{X X} X^2} , 
 \\
 \mathcal{C}^0_3&=\frac{M^2_{pl}}{2\dot{\phi}^{ }K_{XX} X+12HG_{X } X+12HG_{X X} X^2},
 \\
 \mathcal{D}^0&= -\frac{3}{2}K_X + K_{XX} X+6H \dot{\phi}G_{X X}X ,
\\
\mathcal{E}_1^0 &= -\frac{ \dot{\phi} K_X \mathcal{C}^0  + \mathcal{D}^0\left(\mathcal{A}^0\right)^2 + 3M^2_{pl}}{2M^2_{pl}-2G_X X \mathcal{A}^0},
\\
\mathcal{E}_2^0 &= -\frac{  \dot{\phi} K_X \mathcal{C}^0_3 + M_{pl}^2}{4M^2_{pl}-4G_X X \mathcal{A}^0},
\\
\mathcal{E}_3^0 &= \frac13 \mathcal{E}^0_2.
\end{align}
Note that we have assumed $\dot{\phi}^{ }K_{XX} X+6HG_{X } X+6HG_{X X} X^2\neq 0$ and $\mathcal{A}^0 +6M_{pl}^2H/\dot{\phi}K_X \neq 0$, which, by using the background EoMs, is equivalent to $G_X (K_X-K_{XX} X) + K_X G_{XX} X \neq 0$ and $K(G_X K_X-2G_X K_{XX} X + 2K_X G_{XX} X) + K_X^2 G_X  X \neq 0$. So the covariant cubic Galileon case is included in our solution. We will not discuss the special cases where any of the aforementioned quantities actually vanish, but it is easy to follow our formalism in the last section to get the relevant results. The constraint eq. (\ref{ti2order}) now becomes
\begin{equation}
2(G_{X } X\mathcal{A}^0{} - M^2_{pl})\pd_i C^{(1)}_\zeta (\bfx) = M_{pl}^2 \uD^{(1)}_k  C_A^{(1)}{}^{k}_{i}(\bfx).
\end{equation}
Finally, the solution for the late time of inflation is given by
\begin{align}
{\zeta}(t,\bfx)  &= \zeta^{(0)}(\bfx) -\frac{C{}_{\zeta}^{(1)}({\bfx})}{3Ha^3}  + \frac{\mathcal{E}^0_1 \left(C^{(1)}_{\zeta}(\bfx)\right)^2 }{18H^2 a^6}  
 + \frac{\mathcal{E}^0_2 C_A^{(1)}{}^k_l({\bfx})C^{(1)}_A{}^l_k({\bfx}) }{18H^2 a^6}
\nn
&~~~ -  \frac{\mathcal{E}^0_3{}^{[3]}R^{(2)}(\bfx)}{2H^2a^2}  +\mathcal{O}(\epsilon^3),
\\
\label{philtrs}
\phi(t,\bfx) &= \phi^{(0)}(t)+C^{(1)}_\phi(\bfx) - \frac{\mathcal{A}^0 C{}_{\zeta}^{(1)}({\bfx})}{3Ha^3}  +  \mathcal{A}^0 {\zeta}^{(2)}(t,\bfx) + \frac{ \mathcal{C}^0 \left({C}^{(1)}_\zeta(\bfx)\right)^2}{6Ha^6}   
 \nn
 &~~~  - \frac{\mathcal{C}^0_3\,{}^{[3]}R^{(2)}(\bfx)}{4Ha^2}  +\frac{\mathcal{C}^0_3 C^{(1)}_A{}^k_j(\bfx)C^{(1)}_A{}^j_k(\bfx)}{12Ha^6}+\mathcal{O}(\epsilon^3),
\\
A{}^i_{j}(t,{\bfx}) &= \frac{{C}^{(1)}_{A}{}^i_j({\bfx})}{a^3}   +\frac{C^{(1)}_{\zeta}(\bfx){C}^{(1)}_{A}{}^i_j({\bfx})}{Ha^6}
+\frac{{}^{[3]}R^{(2)}{}^i_j(\bfx) -\frac13 \delta^i_j {}^{[3]}R^{(2)}(\bfx)}{Ha^2}+\mathcal{O}(\epsilon^3),
\\
h_{ij}(t,{\bfx}) &= h^{(0)}_{ij}(\bfx) +\frac{2h^{(0)}_{ik}({\bfx}) C^{(1)}_{A}{}^{k}_{j}({\bfx})}{3Ha^3} 
+\frac{2 h^{(0)}_{ik}(\bfx)  \left({}^{[3]}R^{(2)}{}^k_j(\bfx) -\frac13 \delta^k_j {}^{[3]}R^{(2)}(\bfx)\right)}{3 H^2 a^2}
\nn
&~~~ 
 + \frac{3 h^{(0)}_{ik}(\bfx) C^{(1)}_{\zeta}(\bfx){C}^{(1)}_{A}{}^k_j({\bfx}) + 2 h^{(0)}_{il}(\bfx)C_A^{(1)}{}^l_k(\bfx)C_A^{(1)}{}^k_j(\bfx)}{9H^2a^6}
+\mathcal{O}(\epsilon^3).
\end{align} 

Now, we want to explicitly do away with the gauge mode $C^{(1)}_\phi(\bfx)$ in the case of near de Sitter inflation by re-slicing. To this end, we choose 
\begin{align}
\label{timers}
\bar{t} &= t+\eta^0(\bfx) ,
\\
\label{spatialrs}
\bar{x}^i & = x^i + \eta^i  ,
\end{align}
with
\begin{align}
\eta^0(\bfx) &= \frac{C^{(1)}_\phi(\bfx)}{\dot{\phi}^{(0)}} ,
\\
  \eta^i &= \int^t \ud t' \gamma^{ij}({t'},\bfx) \pd_j \eta^0(\bfx) = - \frac{h_{(0)}^{ij}(\bfx)\pd_j C^{(1)}_\phi(\bfx)}{2H\dot{\phi}^{(0)}e^{2\zeta^{(0)}(\bfx)}a(\bar{t})^2} + \mathcal{O}(\epsilon^3).
\end{align}

Let us consider the effects of the temporal transformation on $\phi^{(0)}(t)$: Taylor expansion yields $\phi^{(0)}(t)=\phi^{(0)}(\bar{t})-C^{(1)}_\phi(\bfx)+\mathcal{O}(\epsilon^3)$, which straightforwardly removes the constant mode in eq. (\ref{philtrs}). Though far less obvious, any other effect of the temporal or the spatial part of he transformation leads to corrections that are either $\mathcal{O}(\epsilon^3)$ or can be absorbed in redefinitions of  $C^{(1)}_\zeta(\bfx)$ and $C^{(1)}_A{}^i_j(\bfx)$. The end result is that by re-slicing one can eliminate $C^{(1)}_\phi(\bfx)$ with all the other terms in the solution unchanged.

\section{Conclusions and future perspectives}\label{conclusion}~

In this paper, we have developed the superhorizon gradient expansion formalism for G-inflation, a novel inflation model characterized by its higher order derivative interactions. This model is inspired by a new class of infrared modifications of gravity, called (generalized) Galileon models, introduced to explain the late time accelerated cosmic expansion. There are many interesting features in Galileon inflation, including new shapes of non-Gaussianity \cite{Creminelli:2010qf}.  We have solved the equations of motion of Galileon inflation up to second order in gradient expansion in the synchronous gauge, and obtained the general solution without imposing extra conditions on the first order quantities. We have identified the physical degrees of freedom in the solution, taking particular care in keeping track of the residual gauge freedom left after imposing the synchronous gauge condition. We have also defined a curvature perturbation $\mathcal{R}^{(1)}$ conserved up to first
  order. Finally, we have considered the special case of quasi-de Sitter expansion and we have showed that the general solution is substantially simplified in this case .

Non-Gaussianity in primordial perturbations can be a powerful probe of different inflation models and the associated fundamental theory on which they are based. The gradient expansion technique (valid outside the horizon) is complementary to usual second-order perturbative theory (applied inside the horizon), rather than a complete alternative. In rough terms, one uses usual nonlinear perturbative theory to calculate the generation of non-Gaussianties inside the horizon and uses the gradient expansion to evolve the non-Gaussianities outside the horizon. 
Evolution of non-Gaussianities outside the horizon is often tackled with the separate Universe approximation, which is just the leading order gradient expansion. However, this approximation may be inadequate in some multi-field models or when the slow-roll condition is temporally violated \cite{Leach:2001zf, Naruko:2012fe}, in which case a gradient expansion to second order is needed.

With the formalism developed here, the natural next step  is to calculate non-Gaussiani\-ties in G-inflation at superhorizon scales. In combination 
with the conventional non-linear perturbation analysis inside the horizon, one can then use the existing data to constrain the model parameters (see \cite{Takamizu:2010xy} for an attempt in this direction for k-inflation). Unfortunately, this is not something that can be done straightforwardly  in our case. First of all, there is an important difficulty one has to overcome: after expanding to second order in the gradient expansion, the usual curvature perturbation is not conserved in time and one has to find a new non-linear curvature perturbation. A new curvature perturbation has actually been found in Ref \cite{Takamizu:2013gy} (in uniform Hubble gauge), but under the assumption that the starting order of $\dot{\gamma}_{ij}$ is the second order in the gradient expansion, which largely simplifies the whole calculation. But it is unclear how restrictive this condition is and our results seem to indicate that it is not generically justified. Without this assumption, identifyin
 g the conserved curvature perturbation is a pending, quite non-trivial task. The development of the gradient expansion formalism is only one of the necessary tools for calculating the bispectrum. Some of the other tools already exist (e.g., the second order perturbation analysis inside the horizon). Developing the missing ones and combining everything in order to get the desired result goes beyond the scope of this paper.

Another potential application of the formalism developed here can be to gain general insight on the non-linear behaviour of Galileon fields.  A key feature of Galileon gravity is that it is supposed to give rise to $\mathcal{O}(1)$ corrections to general relativity at large distances and yet satisfies stringent constraints at short distances, such as in the solar system where any modification is typically constrained below $\mathcal{O}(10^{-5})$.  This is achieved due to the high degree of non-linearity of the Galileon derivative interactions and the phenomenon is called the Vainshtein mechanism, originally discovered in massive gravity \cite{Vainshtein:1972sx, Hinterbichler:2011tt}. This mechanism is not easy to see in perturbation theory due to its non-linear nature, and the full non-linear problem is difficult to solve. It would be interesting to use the gradient expansion in order to get a deeper understanding of the behaviour of these non-linear interactions, at least in the regime where it is applicable.

Finally, an interesting extension of this work would be to develop a superhorizon gradient expansion for multi-Galileon inflationary scenarios. Having multiple fields is a typical way to generate non-Gaussianity. Non-Gaussianity in the multi-Galileon model has been discussed \cite{RenauxPetel:2011uk, Fasiello:2013dla}.

\vspace{3cm}

\appendix

\section{Dependence of equations of motion in general covariant scalar-tensor theory}  \label{stsystem}

Consider a general covariant scalar-tensor theory of $\phi$ and $g_{\mu\nu}$, given by the action $S(\phi, g_{\mu\nu})$. The equations of motion for this system are
\begin{align}
\mathcal{E}  & = \frac{1}{\sqrt{-g}}  \frac{\delta S}{\delta \phi} = 0,
\qquad
 \mathcal{E}_{\mu\nu} =\frac{1}{\sqrt{-g}} \frac{\delta S}{\delta g^{\mu\nu}}  =0 ,
\end{align} 
with the variation of the action (modulo boundary terms) is given by
\be
\delta S = \int \ud x^D\sqrt{-g}  \left(  \mathcal{E} \delta \phi + \mathcal{E}_{\mu\nu} \delta g^{\mu\nu} \right).
\ee
Now, we assume this action is covariant, which means it is invariant under the following transformation ($\delta_\xi  x^{\mu} = -\xi^\mu$)
\be
\delta_\xi \phi = \mathcal{L}_{\xi}\phi = \xi^\mu \nabla_\mu \phi , 
\qquad
\delta_\xi g^{\mu\nu} =\mathcal{L}_{\xi} g^{\mu\nu}=  2\nabla^{(\mu} \xi^{\nu)}  .
\ee
That is, we have
\be
\delta_\xi S = \int \ud^D x \sqrt{-g} \left(  \mathcal{E} \cdot \xi^\nu \nabla_\nu \phi + \mathcal{E}_{\mu\nu}\cdot 2\nabla^{\mu} \xi^{\nu}   \right) =0.
\ee
After integration by parts, we get $\int \ud^D x \sqrt{-g}\xi^{\nu} \left(  \mathcal{E}  \nabla_\nu \phi +2 \nabla^{\mu}\mathcal{E}_{\mu\nu}    \right) = 0$. Since $\xi^{\nu}$ is arbitrary, we have
\be
 \mathcal{E}  \nabla_\nu \phi =  2 \nabla^{\mu}\mathcal{E}_{\mu\nu}.
\ee
So if the Einstein equations are satisfied ($\mathcal{E}_{\mu\nu} =0$), the scalar equation of motion is automatically satisfied ($\mathcal{E}=0$).

\section{Gradient expansion of the equations of motion in synchronous gauge} 
\label{sec:gradexpeom}

Here we list the Einstein tensor, the effective energy momentum tensor and the scalar current (the $t$ component) up to order $\mathcal{O}(\epsilon^2)$ in the superhorizon gradient expansion. As in the main text, we suppress the background quantities' order indication ${}^{(0)}$. For example, \emph{$\dot{\phi}^{(0)}$ is written as $\dot{\phi}$}

The quantities at the $\mathcal{O}(\epsilon^0)$ order:
\begin{align} 
G^{(0)}_{tt} &=3H^2,
\\
G^{(0)}_{ti} &=0,
\\
G^{(0)}{}^{i}_{j} &=-\delta^i_j\left(2\dot{H}+3H^2\right)  ,
\\
T^{(0)}_{\phi}{}_{tt} &= -K + 2K_X X +6H\dot{\phi}G_{X } X  ,
\\
T^{(0)}_{\phi}{}_{ti}&= 0   ,
\\
T^{(0)}_{\phi}{}_j^{i} &= K - 2G_{X } X \ddot{\phi}   ,
\\
J^t{}^{(0)} &=K_X^{ }\dot{\phi} +6HG_{X } X   .
\end{align}
The quantities at the $\mathcal{O}(\epsilon)$ order:
\begin{align}
G^{(1)}_{tt} &=6H\dot{\zeta}^{(1)},
\\
G^{(1)}_{ti}&=\frac{2}{3}\uD_i \mathcal{K}^{(0)} = 0,
\\
G^{(1)}{}_j^{i} &=-2\left(\ddot{\zeta}^{(1)}+3H\dot{\zeta}^{(1)}\right)\delta^i_j-\left(3HA^{(1)}{}^i_j+\dot{A}^{(1)}{}^i_j\right)   ,
\\
T^{(1)}_{\phi}{}_{tt} &= \left(  \dot{\phi} K_X + 2\dot{\phi} K_{XX} X + 18HG_X X + 12HG_{XX} X^2 \right) \dot{\phi}^{(1)} + 6\dot{\phi}G_{X } X \dot{\zeta}^{(1)}    ,
\\
T^{(1)}_{\phi}{}_{ti}&= 0   ,
\\
T^{(1)}_{\phi}{}_j^{i} &= \left[\left( \dot{\phi}K_X -  2\dot{X} (G_X + G_{XX} X) \right)\dot{\phi}^{(1)}  -2G_X X\ddot{\phi}^{(1)}  \right] \delta^i_j  ,
\\
J^t{}^{(1)} & = \left( K_X + 2K_{XX}X + 6H\dot{\phi}(G_X + G_{XX} X) \right) \dot{\phi}^{(1)} + 6G_X X \dot{\zeta}^{(1)}  .
\end{align}
The quantities at the $\mathcal{O}(\epsilon^2)$ order:
\begin{align}
G^{(2)}_{tt}&=\frac{1}{2}\left( {}^{[3]}R^{(2)}+6\left(\dot{\zeta}^{(1)}\right)^2+12H\dot{\zeta}^{(2)}-A^{(1)}{}^k_l A^{(1)}{}^l_k\right),
\\
G^{(2)}_{ti}&= -2\uD_i \dot{\zeta}^{(1)} - \uD_k A^{(1)}{}^k_i ,
\\
G^{(2)}{}^{i}_{j}&= {}^{[3]}G^{(2)}{}^i_j
 -\dot{A}^{(2)}{}^{i}_{j}-3HA^{(2)}{}^i_{j}-3\dot{\zeta}^{(1)}A^{(1)}{}^i_{j} 
 \nn
  &~~~ -\left( 2\ddot{\zeta}^{(2)}+6H\dot{\zeta}^{(2)}+3\left(\dot{\zeta}^{(1)}\right)^{2}+\frac{1}{2}A^{(1)}{}^{k}_{l}A^{(1)}{}^{l}_{k}\right)\delta^{i}_j   ,
\\
T^{(2)}_{\phi}{}_{tt} &= (\dot{\phi}K_{X}+2\dot{\phi}K_{XX}X +18 HG_X X + 12 H G_{XX} X^2) \dot{\phi}^{(2)} + 6\dot{\phi}G_X X \dot{\zeta}^{(2)}
\nn
&~~~+\! \left(\!\frac12 K_X \!+\! 4K_{XX}X \!+\! 2 K_{XXX}X^2 \!+\! 9 H\dot{\phi} G_{X} \!+\! 21 H\dot{\phi} G_{XX} X \!+\! 6 H\dot{\phi}G_{XXX}X^2\!\right) \!\! \left(\dot{\phi}^{(1)}\right)^2
\nn
 &~~~ + (18G_X X + 12 G_{XX} X^2) \dot{\zeta}^{(1)} \dot{\phi}^{(1)}  ,
\\
T^{(2)}_{\phi}{}_{ti}&= (\dot{\phi}K_X^{ } + 6HG_{X } X) \pd_i \phi^{(1)} - 2G_X X \pd_i \dot{\phi}^{(1)}  ,
\\
T^{(2)}_{\phi}{}_j^{i} &=  \delta^i_j \left[ -2 G_X X \ddot{\phi}^{(2)} +  \left(\dot{\phi} K_X - 2 \dot{X}(G_X + G_{XX}X) \right) \dot{\phi}^{(2)}  - 2 \dot{\phi} (G_X + G_{XX} X)\dot{\phi}^{(1)}\ddot{\phi}^{(1)} \right.
\nn
 &~~~ \left.  +\left( \frac12 K_X +K_{XX} X - \ddot{\phi} G_X  - 5\ddot{\phi} G_{XX}X - 2\ddot{\phi} G_{XXX}X^2 \right)\left(\dot{\phi}^{(1)}\right)^2  \right]           ,
\\
J^t{}^{(2)} &= 6G_XX\dot{\zeta}^{(2)}+\left(K_X+2K_{XX}X+6H\dot{\phi}G_{X}+6H\dot{\phi}G_{XX}X\right)\dot{\phi}^{(2)} 
\nn
&~~~ +\left(\frac{3}{2}\dot{\phi}K_{XX}+\dot{\phi}K_{XXX}X+3HG_X+15HG_{XX}X+6HG_{XXX}X^2\right)\left(\dot{\phi}^{(1)}\right)^2
\nn
&~~~ +\left (6\dot{\phi}G_X +6\dot{\phi}G_{XX}X \right)\dot{\zeta}^{(1)}\dot{\phi}^{(1)}   .
\end{align}

\acknowledgments

We would like to thank Tsutomu Kobayashi, Arif Mohd,  Shinji Mukohyama, Paul Saffin and Alessanda Silvestri  for helpful discussions. 
The research leading to these results has received funding 
from the European Research Council under the European Union's Seventh Framework Programme (FP7/2007-2013) / ERC Grant Agreement n.~306425 ``Challenging General Relativity'' and from 
the Marie Curie Career Integration Grant LIMITSOFGR-2011-TPS Grant Agreement n.~303537

\end{document}